\documentclass[conference]{IEEEtran}
\IEEEoverridecommandlockouts
\pdfoutput=1
\newcommand{\numRepos}{103\xspace}
\newcommand{\numTotalIssues}{54\,755\xspace}
\newcommand{\numFilteredIssues}{10\,459\xspace}

\newcommand{\numRandomizedSampleSize}{160\xspace}
\newcommand{\numSizeSemanticRandomSample}{119\xspace}
\newcommand{\numSizeMemoryRandomSample}{2\xspace}
\newcommand{\numSizeConcurrencyRandomSample}{4\xspace}
\newcommand{\numRemovedFromRandomizedSampleSet}{35\xspace}

\newcommand{\numKeywordSearchTotal}{756\xspace}
\newcommand{\numKeywordSearchAlreadyInspected}{471\xspace}
\newcommand{\numSizeSemanticKeywordSample}{150\xspace}
\newcommand{\numSizeMemoryKeywordSample}{119\xspace}
\newcommand{\numSizeConcurrencyKeywordSample}{118\xspace}
\newcommand{\numRemovedFromKeywordSampleSet}{84\xspace}

\newcommand{\numSizeSemantic}{269\xspace}

\newcommand{\numSizeTrainingSetConcurrency}{122\xspace}
\newcommand{\numSizeTrainingSetMemory}{121\xspace}
\newcommand{\numSizeTrainingSetSemantic}{126\xspace}
\newcommand{\numSizeTrainingSet}{369\xspace}

\newcommand{\numMlRuns}{100\xspace}

\newcommand{\LSVCmeanFone}{0.72\xspace}

\newcommand{\LSVCmeanPrecision}{0.74\xspace}

\newcommand{\LSVCmeanRecall}{0.72\xspace}

\newcommand{\LSVCmaxRecall}{0.85\xspace}
\newcommand{\LSVCmaxPrecision}{0.88\xspace}
\newcommand{\LSVCmaxFone}{0.85\xspace}

\usepackage{cite}
\usepackage{amsmath,amssymb,amsfonts}
\usepackage{algorithmic}
\usepackage{graphicx}
\usepackage{textcomp}
\usepackage{xcolor}
\def\BibTeX{{\rm B\kern-.05em{\sc i\kern-.025em b}\kern-.08em
    T\kern-.1667em\lower.7ex\hbox{E}\kern-.125emX}}
\usepackage{footmisc}
\usepackage{booktabs} %
\usepackage{tcolorbox}
\usepackage{multirow}
\usepackage[export]{adjustbox}

\usepackage[colorinlistoftodos]{todonotes} %
\presetkeys%
{todonotes}%
{inline,backgroundcolor=yellow}{}

\usepackage{csquotes}
\usepackage{xspace}
\usepackage[bookmarks=false]{hyperref}
\hypersetup{nolinks=false}
\newcommand{\plotPath}{sections/}

\usepackage[top=0.9in ,foot=-0in, bottom=0.8in, left=0.6in, right=0.6in]{geometry}

\usepackage{fancyhdr}
\makeatletter
\patchcmd{\@maketitle}
  {\addvspace{0.5\baselineskip}\egroup}
  {\addvspace{-1\baselineskip}\egroup}
  {}
  {}
\makeatother

\begin{document}

\title{Root cause prediction based on bug reports%
}

\author{\IEEEauthorblockN{Thomas Hirsch}
\IEEEauthorblockA{\textit{Institute of Software Technology} \\
\textit{Graz University of Technology}\\
Graz, Austria \\
thirsch@ist.tugraz.at}
\and
\IEEEauthorblockN{Birgit Hofer}
\IEEEauthorblockA{\textit{Institute of Software Technology} \\
\textit{Graz University of Technology}\\
Graz, Austria \\
bhofer@ist.tugraz.at}
}
\maketitle

\begin{abstract}
This paper proposes a supervised machine learning approach for predicting the root cause of a given bug report.
Knowing the root cause of a bug can help developers in the debugging process---either directly or indirectly by choosing proper tool support for the debugging task.
We mined \numTotalIssues closed bug reports from the issue trackers of \numRepos GitHub projects and applied a set of heuristics to create a benchmark consisting of \numFilteredIssues reports.
A subset was manually classified into three groups  (semantic, memory, and concurrency) based on the bugs' root causes.
Since the types of root cause are not equally distributed, a combination of keyword search and random selection was applied.
Our data set for the machine learning approach consists of \numSizeTrainingSet bug reports (\numSizeTrainingSetConcurrency concurrency, \numSizeTrainingSetMemory memory, and \numSizeTrainingSetSemantic semantic bugs).
The bug reports are used as input to a natural language processing algorithm.
We evaluated the performance of several classifiers for predicting the root causes for the given bug reports.
Linear Support Vector machines achieved the highest mean precision (\LSVCmeanPrecision) and recall (\LSVCmeanRecall) scores.
The created bug data set and classification are publicly available.
\end{abstract}

\footnotetext{\textcopyright 2020 IEEE.  Personal use of this material is permitted.  Permission from IEEE must be obtained for all other uses, in any current or future media, including reprinting/republishing this material for advertising or promotional purposes, creating new collective works, for resale or redistribution to servers or lists, or reuse of any copyrighted component of this work in other works.}

\begin{IEEEkeywords}
bug report, bug benchmark, root cause prediction
\end{IEEEkeywords}
\thispagestyle{fancyplain}

\section{Introduction}\label{sec:intro}

Debugging is one of the most time-consuming parts in the software development process.
While there exist numerous fault localization \cite{WongSurvey2016} and repair \cite{Gazzola2019} techniques to support programmers in the debugging process, it is often unclear which techniques work best for a given bug.
For this reason, Sobreira \textit{et al.}~\cite{Sobreira2018} %
investigated the structure of Defects4J~\cite{Just2014} bugs.
For each bug, they determined the size of the patch, the repair action, and the change pattern. 
They have invited other researchers to investigate which types of bugs\footnote{http://program-repair.org/defects4j-dissection/}
can be handled by their repair tools.

In this paper, we change the perspective of this research topic:
instead of only providing root cause information for a benchmark to help researchers in evaluating their tools, we predict the root cause for a given bug description so that programmers can choose a proper tool for their debugging problem.
There are tools that focus on concurrency (e.g. ConcBugAssist \cite{Khoshnood2015}) or memory (e.g. Valgrind) bugs, while others are better suited for semantic bugs (e.g. Jaguar~\cite{Ribeiro2018}).
While some root causes can easily be determined when reading a bug report,
 other root causes are not that obvious.
Consider for example issue ticket~\#514 from TwelveMonkeys project\footnote{https://github.com/haraldk/TwelveMonkeys/issues/514}:

		\begin{tcolorbox}[boxsep=1pt,left=2pt,right=2pt,top=2pt,bottom=2pt]
\small{
\textbf{TIFF: Invalid StripByteCounts when writing large resolution files (9800x8000)}\\
Hello, when writing a high resolution tiff file the stripByteCounts appears to be corrupt. An approx 300 mb output file has a single image strip with the byte count of: 4071696385 which is larger than the file itself.
However when working with lower (more common) resolutions the meta for the image strips is created properly.
[\dots]
This code creates the file with the incorrect meta data:

\texttt{// Input high resolution 48 bit depth
final InputStream inStream = } [\dots]\\
Attaching zipped image:  9800x8000\_resolution\_48bit\_depth.zip

I've tested and reproduced the issue with the following versions: 3.4.1, 3.4.2, 3.4.3

Thanks in advance,\\
-Jesse
}
\end{tcolorbox}

Our goal is to provide information to the programmer about the root cause of this bug.
For instance, the incorrect byte count mentioned in this bug report together with the information about high resolution can raise suspicion of an integer overflow occurring.

We propose a supervised machine learning (ML) approach that uses the bug description from issue tickets to predict the root cause of the bug.
For processing the text from the issue tickets, we make use of natural language processing (NLP).
For creating the training set, we have mined bug reports from \numRepos GitHub projects and manually examined a subset, classifying them as memory, concurrency or semantic bugs based on the actual fix.
Since the number of concurrency and memory bugs is usually very low \cite{Tan2014a}, we have performed a keyword search in the commit messages of fixes
to find more instances with these root causes.

While the primary goal of this paper is the root cause prediction approach, the generated training data can be used as a benchmark for specific types of faults.
Often, researchers focus on certain bug types when developing a fault localization or repair method.
While these approaches have a high potential, their evaluation is often limited to a few real-world bugs or artificially seeded bugs, as mentioned in~\cite{Tang2008}.
The training data set created in this paper can be used as a bug benchmark by researchers who are interested in certain types of bugs.
It can be seen as a Java pendant to the C/C++ benchmark BugBench that also distinguishes memory, concurrency, and semantic bugs.
Furthermore, it can be used to evaluate information retrieval based bug localization approaches~\cite{Le2017}.

The contributions of this work can be summarized as:
\begin{itemize}
	\item a machine learning approach for predicting the root cause for a given bug report with a mean precision of \LSVCmeanPrecision and a mean recall of \LSVCmeanRecall,
	\item a data set consisting of \numFilteredIssues bug reports and fixes from \numRepos GitHub repositories,
	\item a data set of \numSizeTrainingSetConcurrency concurrency, \numSizeTrainingSetMemory memory, and \numSizeSemantic semantic bugs with detailed sub-categories, and
	\item a framework for building such data sets.
\end{itemize}

The created data sets, all scripts, and the categorization are publicly available.\footnote{https://doi.org/10.5281/zenodo.3973048\label{footnote:onlineAppendix}}
The structure of this paper is as follows:
Section~\ref{sec:schema} introduces the main root cause categories and their sub-categories.
Section~\ref{sec:dataAquisition}  explains how we have collected closed bug reports and their corresponding fixes.
Section~\ref{sec:ml} presents the machine learning approach.
We discuss the results and threats to validity in Section~\ref{sec:results}.
Section~\ref{sec:rw} discusses the related work and Section~\ref{sec:conclusion} concludes the paper.

\section{Classification schema}\label{sec:schema}
We use three main categories %
and 18~detailed root causes as described in Table~\ref{tab:CategoritzationRootCause}.
The semantic and memory sub-categories are based on Tan \textit{et al.}~\cite{Tan2014a};
the concurrency sub-categories are based on Zhou \textit{et al.}~\cite{zhou2015}.

\begin{table}[htbp]
	\centering
		\caption{Root cause categories and sub-categories}
	\label{tab:CategoritzationRootCause}

		\begin{tabular}{lp{6cm}}
\toprule
\textbf{Semantic} & \textbf{Description} \\ \midrule
			
 \textit{Exception handl.} 
 &	Missing or improper exception handling. %
\\
\textit{Missing case} 
 & Faults due to unawareness of a certain case or simply a forgotten implementation.\\
\textit{Processing}
& Incorrect implementation (e.g.  miscalculations, incorrect method output, wrong method/library usage).\\
 \textit{Typo}
& Ambiguous naming, typos in SQL calls/URLs/paths. \\
 \textit{Dependency}
& The code can be built but behaves unexpected because of changes in a foreign system (e.g. update of utilized library or underlying OS). 
	\\
\textit{Other}
& All other semantic faults. %
\\

\toprule
\textbf{Memory} &  \\ \midrule %
					 \textit{Buffer overflow} & Buffer overflows, not overflowing numeric types.\\
				 \textit{Null pointer deref.}& All null pointer dereferences.\\
		 \textit{Uninit. mem. read}&  All uninitialized memory reads except null pointer dereference.  \\
				 \textit{Memory leak}& Memory leak.\\
					 \textit{Dangling pointer}& Dangling pointer.\\
				 \textit{Double free}	& Double free.\\
			 \textit{Other}& All other memory bugs.\\

\toprule
\textbf{Concurrency} & \\ \midrule %
\textit{Order violation} 
& Missing or incorrect synchronization, e.g. object is dereferenced by thread B before it is initialized by thread~A.\\
 \textit{Race condition} 
& Two or more threads access the same resource with at least one being a write access and the access is not ordered properly.\\
 \textit{Atomic violation} 
& Constraints on the interleaving of operations are missing. This happens when atomicity of a certain code region was assumed but failed to guarantee atomicity in the implementation. 
\\
	\textit{Deadlock}
&  Two or more threads wait for the other one to release a resource. \\
  \textit{Other} 
& All other concurrency bugs. \\ \midrule
		\end{tabular}
\end{table}

A problem with post mortem bug classification arises through often unclear separation of the actual fix from other code changes, e.g., commits that include more than one bug fix, commits that include the bug fix aside of some refactoring or new extension, or bug fixes that are scattered over multiple commits.
Additionally, it is difficult to distinguish a workaround from a fix \cite{Bohme2017}.
All of the above make it hard to correctly identify the fix and to properly categorize the root cause. 
To deal with these issues, we have added a confidence value ranging from 1-10 that reflects our confidence on the correctness of our classification:
A confidence level of 10 indicates showcase quality; 
9 indicates that we are very confident about the main category and the subcategory;
8 indicates that we are very confident about main category and subcategory assigned, but a different subcategory cannot be ruled out with 100 \% certainty.
For example, differentiating \enquote{processing} and \enquote{missing case} is often not possible without having the knowledge of the programmer who wrote the code.
A confidence level of 7 or below indicates doubts about the chosen subcategory.
Confidence levels between 3 and 5 indicate a strong confidence about the main category, but the subcategories were not identifiable.
A confidence level of 2 indicates doubts about the main category while
a level of 1 indicates that it was not possible to determine the main root cause category for the bug.
	
\section{Data acquisition}\label{sec:dataAquisition}
In this section, we provide details on the collection of the bug data set that builds the basis for creating the training set for the machine learning approach.

\emph{Purpose of the data set.}
The data set should provide a realistic distribution of different bug types, and should serve as basis for experiments with various fault localization and ML experiments.
The bugs should be real world Java bugs.

\emph{Project selection.}
\label{emph:projectSelection}
We chose \numRepos GitHub Java projects to source our data set.
Primary selection criteria were a well known organization driving the project, or the project having a high star rating on GitHub.
However, the list also contains lesser known projects that were already used in other research \cite{Gyimesi2015a, Toth2016, Just2014}.
The selection process was performed manually.
All of the projects utilize GitHub's built-in issue tracker together with its labeling system, and have at least 100 closed issues identified as bugs.
The project sizes range from 13k~Java LOC (Lines Of Code) for HikariCP\footnote{https://github.com/brettwooldridge/HikariCP} to 1.7M Java LOC for Elasticsearch\footnote{https://github.com/elastic/elasticsearch}.
The full list of mined projects can be found in the online appendix\footref{footnote:onlineAppendix}.

\emph{Bug ticket identification.}
We identified bugs via the labels used in the issue tickets and we only considered closed issue tickets.
In order to omit feature requests, maintenance tickets and other non-bug issues, we only considered issues whose labels contain \enquote{bug}, \enquote{defect}, or \enquote{regression}.%

\emph{Filtering criteria.}
GitHub automatically links commits to  issue tickets based on issue ids and provides this data together with issue tickets.
We only consider issues for which at least one commit is linked to the issue, and all linked commits are still available in the Git repository.
If any of the commits are linked to multiple issues, or the commit message suggests that the fix is done aside of other changes, the issue is discarded.
As of writing this, we omit issues whose commits do not contain changes in production Java code. %
We plan to lift this limitation to incorporate other root causes, e.g. in the documentation or build system in the future.

We use Gumtree Spoon AST Diff\footnote{https://github.com/SpoonLabs/gumtree-spoon-ast-diff}\cite{Falleri2014} to create Java aware diffs.
To manage overwhelming runtime and memory requirements arising from the size of the data set, we limit the size and number of the commits per issue.
We only consider issues where the number of commits linked to the issue is smaller than 10, the number of files changed per commit is smaller than 20, and the number of lines changed per commit is smaller than 250. %
Our analysis shows that these limitations only remove less than 3\,\% of the issues. %

\emph{The data set.}
In total, \numTotalIssues issues have been mined from GitHub.
Following the filtering criteria described above leaves us with \numFilteredIssues issues that form the basis for our further investigations.
This bug data set consists of:
\begin{itemize}
	\item textual bug report including metadata in form of time-stamps and user names,
	\item all commits associated to each bug report including metadata as commit message and git stats, and
	\item Java aware diff statistics and location of the changes in terms of file, class, and method.
\end{itemize}

\section{Machine learning approach}\label{sec:ml}
We employ an NLP approach, vectorizing the textual bug reports into unigrams and bigrams, to train a model for automated classification along our fault classification schema.
This approach calculates a frequency vector from words and word pairs occurring in the input text that is used as feature vector for the classifier.

\emph{Input preprocessing.}
To increase performance  of the classifier, we applied the following preprocessing steps: %
\begin{itemize}
	\item Stop word removal (i.e. removing common words that are not adding any value)
	\item Case folding (i.e. converting all characters to lower case)
	\item Stemming (i.e. reducing each word to its word stem)
\end{itemize}
The bug reports often include stack traces, exceptions, and log outputs.
Currently, we process them in the same way as the rest of the input text.
In future work, we will investigate the usefulness of domain specific preprocessing of these artifacts.

\begin{figure}[t]
	\centering                   %
		\includegraphics[width=\columnwidth, trim={7.7cm 1.0cm 7.7cm 0.2cm},clip]{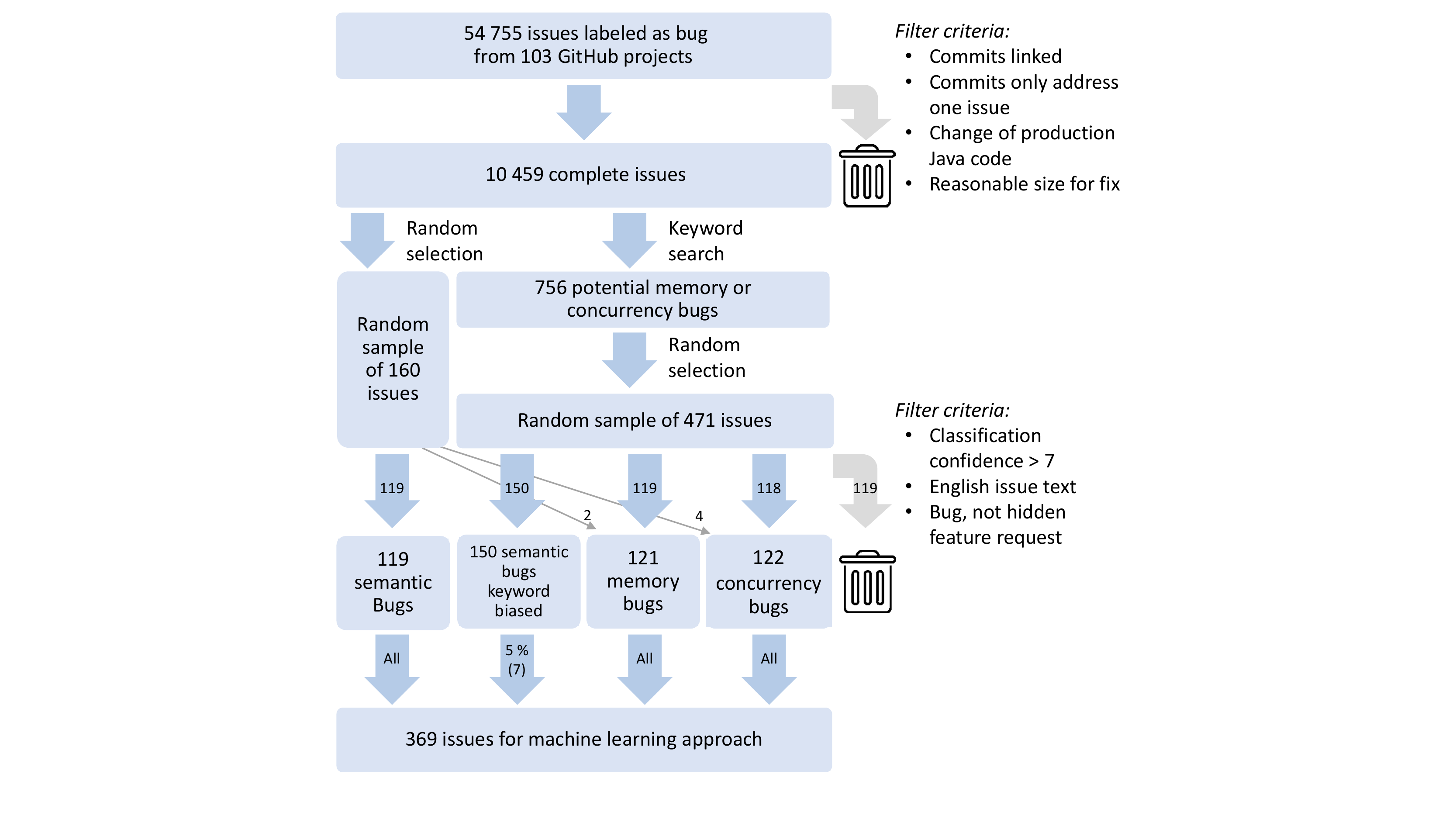}
	\caption{Training set creation}
	\label{fig:trainingSetCreation}
\end{figure}

\emph{Training set creation.}
Figure~\ref{fig:trainingSetCreation} provides an overview of the training set creation. 
We manually classified  \numRandomizedSampleSize randomly selected issues and identified \numSizeSemanticRandomSample semantic, \numSizeMemoryRandomSample memory, and \numSizeConcurrencyRandomSample concurrency bugs.
\numRemovedFromRandomizedSampleSet issues were not classified because the bug reports were non-English, feature requests, deemed not a bug, 
 or issues for which we were not confident about the sub-category (confidence level $<$\,8).

Concurrency and memory bugs are usually rare, accounting for 2\,\% respectively  6\,\% of all bugs \cite{Ray2014,Li2006a, Tan2014a},
which poses a challenge for the creation of reasonably sized training sets.
For this reason, we have performed a keyword search on the commit messages linked to the issues to identify candidates of memory and concurrency bugs analog to Ray \textit{et al.}'s approach\cite{Ray2014}, resulting in 
a total of \numKeywordSearchTotal issues. %
As of writing this, \numKeywordSearchAlreadyInspected randomly selected issues from this set have been examined and classified.
\numSizeSemanticKeywordSample semantic, \numSizeMemoryKeywordSample memory, and \numSizeConcurrencyKeywordSample concurrency bugs have been identified in  this sample.
\numRemovedFromKeywordSampleSet issues could not be classified due to the reasons mentioned above.

\emph{Training set composition.}
To avoid a bias towards the semantic bugs that were \enquote{accidentally found} during the  manual classification of the keyword search results and to have approximately equally large training sets,
we reduced their volume to 5\,\% of all semantic bugs.
This is a rather high estimate given the fact, that only 7.2\,\% of all bugs have been reported in the keyword search and only one third of these bugs are actually semantic bugs.
Further, using separate data bases for the keyword search (commit messages) and training set for our ML classifier (bug reports) makes us confident that the bias introduced by the keywords is limited.
As of writing this, our training set consists of \numSizeTrainingSetConcurrency concurrency bugs, \numSizeTrainingSetMemory memory bugs, and \numSizeTrainingSetSemantic semantic bugs.
The complete training set consists of \numSizeTrainingSet textual bug reports.

\emph{Classifiers.}
We applied various supervised ML classifier algorithms on our data set, namely Multinomial Naive Bayes (MNB), Linear Support Vector (LSVC), Linear Support Vector with Stochastic Gradient Descent learning (SGDC), Random Forrest (RFC), and Logistic Regression (LRC).
The selection of classifiers is based on their suitability for multi-class classification problems based textual inputs, and their application in similar research.
Support vector machines have been used in comparable endeavors\cite{Ortu2016, Thung2012, Ray2014, Li2006a, Tan2014a}; the same applies to naive Bayes\cite{Ortu2016, Hernandez-Gonzalez2018, Antoniol2008, Chawla2015, Li2006a, Tan2014a}, logistic regression\cite{Antoniol2008, Chawla2015}, and decision tree based algorithms\cite{Antoniol2008, Chawla2015, Tan2014a}.

\emph{Experiment.}
The \numSizeTrainingSet bug reports were split into a training set (80\,\%) and a test set (20\,\%).
We performed 5-fold cross validation on the training set for each classifier, using grid search for hyperparameter tuning.
Mean accuracy was used as scoring metric in the grid search.
The highest scoring model for each classifier was then evaluated on the test set yielding the final score for this classifier.
This experiment was performed 10 times, followed by manual examination of its results and corresponding hyperparameters to reduce the grid search space.
Finally, to enable comparison of the classifiers given the small input data set, the above described experiment with the reduced hyperparameter set was performed \numMlRuns times with randomized test and training splits.
The employed set of hyperparameters, and grid search results for each experiment, can be found in the online appendix. %

\section{Results}\label{sec:results}

\subsection{Classification results}
\begin{figure}[htbp]
	\centering
		\includegraphics[width=1.00\columnwidth, trim={0.7cm 0.7cm 1.3cm 1.7cm},clip]{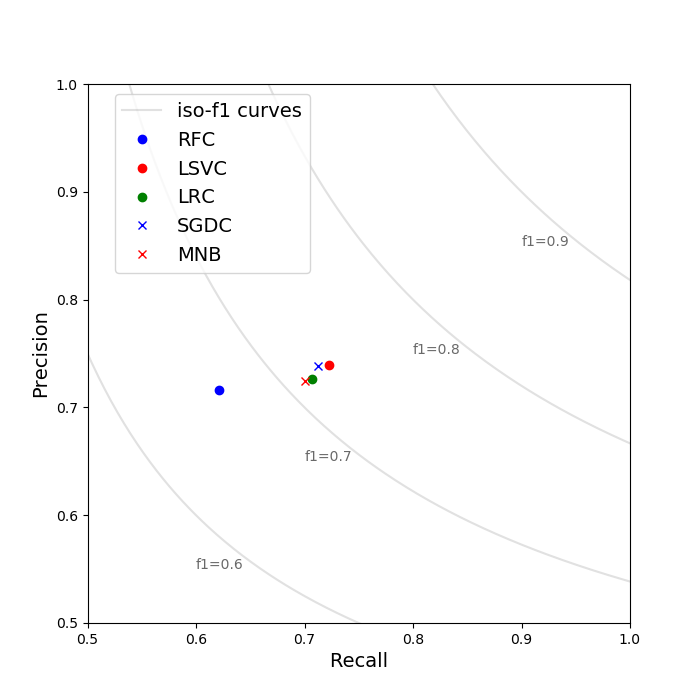}
	\caption{Mean weighted average precision, recall and F1 score}
	\label{fig:isof1}
\end{figure}

We graphically compare the classifiers' performance by means of the weighted averages of F1, precision, and recall in Figure~\ref{fig:isof1} and we report  
mean, median,  standard deviation, min, and max of each classifier in Table~\ref{tab:meanMediaF1PrecisionRecall} based on the scores of \numMlRuns runs.
Please note that the F1 scores are computed for the individual test runs and then the mean, median, standard deviation, min, and max values of these F1 scores are computed.
Thus, they cannot be computed from the precision and recall given in the table. 

We observed a tight clustering of classifiers, which is also evident in individual runs, although individual runs exhibit varying performances.
We attribute this behavior to the small data set size and high variance in data quality.
The best overall performance was achieved with LSVC, with mean  F1 (\LSVCmeanFone%
), precision (\LSVCmeanPrecision%
), and recall (\LSVCmeanRecall %
).
LSVC also produced the highest observed scores in an individual run, yielding F1 (\LSVCmaxFone), precision (\LSVCmaxPrecision), and recall (\LSVCmaxRecall).

\begin{table}
\centering
\caption{Weighted average precision, recall, and F1 scores (\numMlRuns runs)}
	\label{tab:meanMediaF1PrecisionRecall}
\begin{tabular}{llccc}
\toprule
                           & Classifier & Precision & Recall & F1   \\
\midrule
\multirow{5}{*}{Mean}      & LRC        & 0.73         & 0.71      & 0.71   \\
                           & RFC        & 0.72         & 0.62      & 0.62   \\
                           & SGDC       & 0.74         & 0.71      & 0.71   \\
                           & \textbf{LSVC}& \textbf{0.74}& \textbf{0.72}& \textbf{0.72}\\
                           & MNB        & 0.72         & 0.70      & 0.70   \\
\midrule
\multirow{5}{*}{Median}    & LRC        & 0.73         & 0.70      & 0.70   \\
                           & RFC        & 0.72         & 0.62      & 0.62   \\
                           & SGDC       & \textbf{0.74}& 0.70      & 0.71   \\
                           & \textbf{LSVC}& 0.74         & \textbf{0.72}& \textbf{0.72}   \\
                           & MNB        & 0.73         & 0.70      & 0.70   \\
\midrule
\multirow{5}{*}{Std. dev.} & LRC        & 0.046        & 0.048     & 0.048  \\
                           & RFC        & 0.053        & 0.076     & 0.082  \\
                           & SGDC       & 0.045        & 0.049     & 0.049  \\
                           & \textbf{LSVC}& 0.046        & \textbf{0.046}& \textbf{0.046}  \\
                           & MNB        & \textbf{0.045}& 0.049     & 0.049  \\
\midrule
\multirow{5}{*}{Min}       & LRC        & 0.63         & 0.62      & 0.62   \\
                           & RFC        & 0.53         & 0.39      & 0.38   \\
                           & SGDC       & 0.64         & 0.62      & 0.63   \\
                           & \textbf{LSVC}& \textbf{0.65}& \textbf{0.64}& \textbf{0.63}   \\
                           & MNB        & 0.60         & 0.57      & 0.57   \\
\midrule
\multirow{5}{*}{Max}       & LRC        & 0.84         & 0.82      & 0.82   \\
                           & RFC        & 0.85         & 0.77      & 0.77   \\
                           & SGDC       & 0.86         & 0.84      & 0.84   \\
                           & \textbf{LSVC}& \textbf{0.88}& \textbf{0.85}& \textbf{0.85}   \\
                           & MNB        & 0.83         & 0.82      & 0.82   \\
\bottomrule
\end{tabular}
\end{table}

\subsection{Discussion}%
The biggest challenge lies in the creation of a reasonably sized data set. %
Further, varying data quality constitutes a significant problem.
The textual bug reports in our data set range from only 5 words to 60kB of text per report.
However, our examination of those bug tickets shows that the length is not necessarily correlating with the quality in terms of usefulness for the developer.
Issue \#4960 from Elasticsearch\footnote{https://github.com/elastic/elasticsearch/issues/4960} is an example  of a bug report that requires context and knowledge about the project for understanding:
\begin{tcolorbox}[boxsep=1pt,left=2pt,right=2pt,top=2pt,bottom=2pt]
\small{
\textbf{Filtered query parses \_name incorrectly
}}
\end{tcolorbox}

There are bug reports  that merely describe the impact, e.g. issue \#338 from the Redisson project\footnote{https://github.com/redisson/redisson/issues/338}:
\begin{tcolorbox}[boxsep=1pt,left=2pt,right=2pt,top=2pt,bottom=2pt]
\small{
\textbf{New version 2.1.4 or greater performance is low.}

When I use redisson 2.1.3 the ubuntu's load average is 1.8 ~ 2.3; but I use 2.1.4 or greater, the load average is often greater than 3.00, my java application often overload.
}
\end{tcolorbox}

In some cases, the bug reports point right away at the fault, e.g. 
Netty issue \#1878\footnote{https://github.com/netty/netty/issues/1878}:
\begin{tcolorbox}[boxsep=1pt,left=2pt,right=2pt,top=2pt,bottom=2pt]
\small{
\textbf{WebSocket08FrameDecoder leaks ByteBuf when payload is masked}
}
\end{tcolorbox}

Further research is required to determine metrics to measure bug report quality for our purpose.
On the other end of the spectrum, for very long bug reports, additional text pre-processing is required.
Heuristics for reduction or removal of artifacts have to be implemented.
Such artifacts are stack traces, code snippets, log outputs, or similar text portions, whose size is disproportionate to the added information.

\subsection{Threats to validity}
The selection of bugs from the issue tickets by searching for certain labels is a threat to the internal validity.
While we have considered a wide range of bug labels, we cannot rule out to miss bugs with special labels or wrongly labeled bugs.
A study on 7000 issue reports from five open-source projects showed that up to 40\,\% of the issues were wrongly labeled~\cite{Herzig2013}.

Manually categorizing the root cause might be error-prone and the true root cause of the bug can only be determined by the original programmer.
For this reason, we indicated the confidence level for each bug we categorized and excluded bugs with a low confidence level.
Furthermore, the fix might be a workaround instead of a fix of the true fault.

The keyword search might only reveal certain types of memory and concurrency bugs.
We have tried to avoid a bias in the classification towards the words used in the keyword search by performing the keyword search on the commit messages and NLP for classification on the bug description.

The small sample size is the biggest threat to external validity.
In future work, we will therefore enlarge the training set.
The performance of this approach may vary based on the software domain of the examined project.
We tried to counteract this by including software projects to source our data set.
However, data mining was exclusively performed on open source projects.
Further, most of the examined projects are libraries.
In contrast to end-user software, bug reports for libraries are almost exclusively written by other developers.
Such bug reports often already contain insights into the underlying problem.
Further, our approach may not work as good for bug descriptions of software written in a different programming language.

\section{Related Work}\label{sec:rw}
Ray \textit{et al.}~\cite{Ray2014} analyzed more 560\,000 bug fixes from 729~GitHub projects written in 17~languages.
They classified the root causes and impacts for 10\,\% of the bugs by searching for keywords %
in the commit messages and  trained a supervised ML approach to classify the remaining 90\,\% of the bugs. %
They validated their approach by manual classifying 180~bug fixes (83.7\,\% precision,  84.3\,\% recall). 
While we also rely on a keyword search, we did not perform the keyword search on the same text that was used in NLP to avoid biasing.

Li and colleagues~\cite{Li2006a} classified the root cause, impact, and software component of nearly 30\,000 Bugzilla entries using NLP with
SVM, Winnow, Perceptron and Naive Bayes as classifiers.
Their training set consists of 709 bugs (51\,\%  randomly sampled, 36\,\%  security-related, and 13\,\%  concurrency bugs).

Tan \textit{et al.}~\cite{Tan2014a} manually classified 339 bugs from randomly selected fixed issues of three open-source projects %
into the dimensions root cause, impact, and component. 
Because of the low number of concurrency bugs in the sample, they performed a keyword search %
to identify additional concurrency bugs.
Semantic bugs are the dominant root cause with 70-87\,\%.
The Linux kernel has nearly 13.6\,\% concurrency bugs; the other projects (Mozilla and Apache) have a lower number of concurrency bugs with 1.2\,\% and 5.2\,\%.
Furthermore, the authors automatically classified more than 100\,000 bugs using a supervised ML (precision: 67\,\% for memory and 93\,\% for semantic bugs, recall: 57\,\% resp. 95\,\%).

Ortu \textit{et al.}~\cite{Ortu2016} investigated whether there are differences in the characteristics of high and low priority defects in more than 1200 open-source software projects. 
Therefore, they trained different supervised machine learning classifiers to predict the root cause, impact, and software component.

Thung \textit{et al.}~\cite{Thung2012} used machine learning to classify bugs according to the Orthogonal Defect Classification (ODC) scheme.
They distinguished three defect groups: data and control flow, structural, and non-functional.
They manually classified 500 bugs that serve as training set.
They use the description of the bug as well as the fixes to train a model.
The SVM multi-class classification algorithm  performed best (69\,\% precision,  70\,\% recall).
Lopes and colleagues~\cite{Lopes2020} applied different ML algorithms on bug descriptions to classify bugs according to different ODC dimensions.
They manually categorized more than 4000 fixed bugs from three NoSQL databases. %
Recurrent Neural Networks have the highest accuracy when predicting the activity (47.6\,\%) and impact (33.3\,\%).
Linear support vector machines are suited best to predict the target (accuracy 85.5\,\%) and the defect type (34.7\,\%).

Hern\'andez-Gonz\'alez and colleagues~\cite{Hernandez-Gonzalez2018} proposed a learning from crowds ML approach.
The training data consists of bug reports and labels for OCD's impact dimension.
Each bug report was labeled by five annotators.
In the majority of the cases, the annotators disagree on the labels.
In the learning from crowds paradigm, the individual labels are taken in the machine learning training instead of the label that was assigned by the majority of the annotators.

Antoniol \textit{et al.}~\cite{Antoniol2008} use decision trees, naive Bayes and logistic regression to classify issue reports as bug or feature request.
Their approach was able to correctly classify 77-82\,\% of the issues. %
Chawla and Singh~\cite{Chawla2015} also classify issue reports as bug or other request.
They receive an accuracy of 84-91\,\%. %

\section{Conclusion and Future Work}\label{sec:conclusion}
The presented approach automatically predicts the root cause for a given bug report.
This information can be used  by the developer to choose a proper debugging tool.
It can also be used by a meta-debugging approach to recommend a debugging tool.
The data set created in this work can be used to evaluate which debugging tools are particularly well-suited to support programmers in the debugging process of a particular bug instance.
In addition, the proposed approach can be utilized for building benchmarks of specific bug types.
This benchmark is especially suitable for evaluating IR-based fault localization techniques since it includes textual data in form of bug reports and commit messages, as well as detailed information on the fix location.

During the manual classification, we have noticed recurring fault patterns.
We will investigate if we can establish links between these fault patterns  and code-smells detected by existing code analysis tools such as SonarQube\footnote{https://www.sonarqube.org/}.
If so, knowledge about the bug type combined with reports from code analysis tools can be utilized to aid fault localization.

Besides that, we will improve the approach by pre-processing stack trace information and other artifacts (if available in the bug report).
Currently, stack trace information is treated the same way as human written text.

Since a detailed predicted root cause is even more helpful, we will refine the prediction to the sub-categories.
To do so, we have to enlarge the training set.
Since certain subcategories will be underrepresented in the training set, we will up-sample those categories by means of the Synthetic Minority Over-sampling TEchnique (SMOTE)~\cite{Chawla2002}.

\section*{Acknowledgment}
The work described in this paper has been funded by the Austrian Science Fund (FWF): P 32653 (Automated Debugging in Use).

\bibliographystyle{IEEEtran}
\bibliography{library}

\begin{thebibliography}{10}
\providecommand{\url}[1]{#1}
\csname url@samestyle\endcsname
\providecommand{\newblock}{\relax}
\providecommand{\bibinfo}[2]{#2}
\providecommand{\BIBentrySTDinterwordspacing}{\spaceskip=0pt\relax}
\providecommand{\BIBentryALTinterwordstretchfactor}{4}
\providecommand{\BIBentryALTinterwordspacing}{\spaceskip=\fontdimen2\font plus
\BIBentryALTinterwordstretchfactor\fontdimen3\font minus
  \fontdimen4\font\relax}
\providecommand{\BIBforeignlanguage}[2]{{%
\expandafter\ifx\csname l@#1\endcsname\relax
\typeout{** WARNING: IEEEtran.bst: No hyphenation pattern has been}%
\typeout{** loaded for the language `#1'. Using the pattern for}%
\typeout{** the default language instead.}%
\else
\language=\csname l@#1\endcsname
\fi
#2}}
\providecommand{\BIBdecl}{\relax}
\BIBdecl

\bibitem{WongSurvey2016}
W.~E. Wong, R.~Gao, Y.~Li, R.~Abreu, and F.~Wotawa, ``{A Survey on Software
  Fault Localization},'' \emph{IEEE Transactions on Software Engineering},
  vol.~42, no.~8, pp. 707--740, aug 2016.

\bibitem{Gazzola2019}
L.~Gazzola, D.~Micucci, and L.~Mariani, ``{Automatic Software Repair: A
  Survey},'' \emph{IEEE Transactions on Software Engineering}, vol.~45, no.~1,
  pp. 34--67, jan 2019.

\bibitem{Sobreira2018}
\BIBentryALTinterwordspacing
V.~Sobreira, T.~Durieux, F.~Madeiral, M.~Monperrus, and M.~A. Maia,
  ``{Dissection of a Bug Dataset: Anatomy of 395 Patches from Defects4J},''
  \emph{25th IEEE International Conference on Software Analysis, Evolution and
  Reengineering (SANER 2018)}, vol. 2018-March, pp. 130--140, jan 2018.
  [Online]. Available: \url{http://dx.doi.org/10.1109/SANER.2018.8330203}
\BIBentrySTDinterwordspacing

\bibitem{Just2014}
\BIBentryALTinterwordspacing
R.~Just, D.~Jalali, and M.~D. Ernst, ``{Defects4J: a database of existing
  faults to enable controlled testing studies for Java programs},'' in
  \emph{International Symposium on Software Testing and Analysis (ISSTA
  2014)}.\hskip 1em plus 0.5em minus 0.4em\relax ACM Press, jul 2014, pp.
  437--440. [Online]. Available:
  \url{http://dl.acm.org/citation.cfm?doid=2610384.2628055}
\BIBentrySTDinterwordspacing

\bibitem{Khoshnood2015}
S.~Khoshnood, M.~Kusano, and C.~Wang, ``{ConcBugAssist: Constraint solving for
  diagnosis and repair of concurrency bugs},'' in \emph{Int. Symp. on Software
  Testing and Analysis (ISSTA 2015)}.\hskip 1em plus 0.5em minus 0.4em\relax
  ACM, 2015, pp. 165--176.

\bibitem{Ribeiro2018}
H.~L. Ribeiro, H.~A. {De Souza}, R.~P.~A. {De Araujo}, M.~L. Chaim, and F.~Kon,
  ``{Jaguar: A Spectrum-Based Fault Localization Tool for Real-World
  Software},'' in \emph{11th International Conference on Software Testing,
  Verification and Validation (ICST 2018)}.\hskip 1em plus 0.5em minus
  0.4em\relax IEEE, may 2018, pp. 404--409.

\bibitem{Tan2014a}
L.~Tan, C.~Liu, Z.~Li, X.~Wang, Y.~Zhou, and C.~Zhai, ``{Bug characteristics in
  open source software},'' \emph{Empirical Software Engineering}, vol.~19,
  no.~6, pp. 1665--1705, oct 2014.

\bibitem{Tang2008}
Y.~Tang, Q.~Gao, and F.~Qin, ``{LeakSurvivor: Towards Safely Tolerating Memory
  Leaks for Garbage-Collected Languages},'' in \emph{USENIX Annual Technical
  conference}, 2008, pp. 307--320.

\bibitem{Le2017}
T.~D.~B. Le, F.~Thung, and D.~Lo, ``{Will this localization tool be effective
  for this bug? Mitigating the impact of unreliability of information retrieval
  based bug localization tools},'' \emph{Empirical Software Engineering},
  vol.~22, no.~4, pp. 2237--2279, aug 2017.

\bibitem{zhou2015}
\BIBentryALTinterwordspacing
B.~Zhou, I.~Neamtiu, and R.~Gupta, ``{Predicting concurrency bugs: How many,
  what kind and where are they?}'' in \emph{9th International Conference on
  Evaluation and Assessment in Software Engineering (EASE'15)}.\hskip 1em plus
  0.5em minus 0.4em\relax ACM, apr 2015, pp. 1--10. [Online]. Available:
  \url{https://doi.org/10.1145/2745802.2745807}
\BIBentrySTDinterwordspacing

\bibitem{Bohme2017}
\BIBentryALTinterwordspacing
M.~B{\"{o}}hme, E.~O. Soremekun, S.~Chattopadhyay, E.~Ugherughe, and A.~Zeller,
  ``{Where is the bug and how is it fixed? An experiment with practitioners},''
  in \emph{11th Joint Meeting on Foundations of Software Engineering (ESEC/FSE
  2017)}, vol. Part F1301.\hskip 1em plus 0.5em minus 0.4em\relax Association
  for Computing Machinery, aug 2017, pp. 117--128. [Online]. Available:
  \url{http://dl.acm.org/citation.cfm?doid=3106237.3106255}
\BIBentrySTDinterwordspacing

\bibitem{Gyimesi2015a}
P.~Gyimesi, G.~Gyimesi, Z.~T{\'{o}}th, and R.~Ferenc, ``{Characterization of
  source code defects by data mining conducted on GitHub},'' in \emph{Lecture
  Notes in Computer Science}, vol. 9159.\hskip 1em plus 0.5em minus 0.4em\relax
  Springer, 2015, pp. 47--62.

\bibitem{Toth2016}
Z.~T{\'{o}}th, P.~Gyimesi, and R.~Ferenc, ``{A public bug database of GitHub
  projects and its application in bug prediction},'' in \emph{16th Int.
  Conference on Computational Science and Its Applications (ICCSA'16)}, vol.
  9789.\hskip 1em plus 0.5em minus 0.4em\relax Lecture Notes in Computer
  Science, Springer, 2016, pp. 625--638.

\bibitem{Falleri2014}
\BIBentryALTinterwordspacing
J.~R. Falleri, F.~Morandat, X.~Blanc, M.~Martinez, and M.~Monperrus,
  ``{Fine-grained and accurate source code differencing},'' in \emph{29th
  ACM/IEEE International Conference on Automated Software Engineering (ASE
  2014)}.\hskip 1em plus 0.5em minus 0.4em\relax ACM, 2014, pp. 313--323.
  [Online]. Available:
  \url{http://dl.acm.org/citation.cfm?doid=2642937.2642982}
\BIBentrySTDinterwordspacing

\bibitem{Ray2014}
\BIBentryALTinterwordspacing
B.~Ray, D.~Posnett, V.~Filkov, and P.~Devanbu, ``{A large scale study of
  programming languages and code quality in GitHub},'' in \emph{ACM SIGSOFT
  Symposium on the Foundations of Software Engineering (FSE'14)}.\hskip 1em
  plus 0.5em minus 0.4em\relax ACM, nov 2014, pp. 155--165. [Online].
  Available: \url{http://dl.acm.org/citation.cfm?doid=2635868.2635922}
\BIBentrySTDinterwordspacing

\bibitem{Li2006a}
Z.~Li, L.~Tan, X.~Wang, S.~Lu, Y.~Zhou, and C.~Zhai, ``{Have things changed
  now?: An empirical study of bug characteristics in modern open source
  software},'' in \emph{1st Workshop on Architectural and System Support for
  Improving Software Dependability (ASID'06)}, 2006, pp. 25--33.

\bibitem{Ortu2016}
\BIBentryALTinterwordspacing
M.~Ortu, G.~Destefanis, S.~Swift, and M.~Marchesi, ``{Measuring high and low
  priority defects on traditional and mobile open source software},'' in
  \emph{7th International Workshop on Emerging Trends in Software Metrics
  (WETSoM 2016)}.\hskip 1em plus 0.5em minus 0.4em\relax ACM, may 2016, pp.
  1--7. [Online]. Available:
  \url{http://dl.acm.org/citation.cfm?doid=2897695.2897696}
\BIBentrySTDinterwordspacing

\bibitem{Thung2012}
F.~Thung, D.~Lo, and L.~Jiang, ``{Automatic defect categorization},'' in
  \emph{Working Conf. on Reverse Engineering (WCRE)}, 2012, pp. 205--214.

\bibitem{Hernandez-Gonzalez2018}
J.~Hern{\'{a}}ndez-Gonz{\'{a}}lez, D.~Rodriguez, I.~Inza, R.~Harrison, and
  J.~A. Lozano, ``{Learning to classify software defects from crowds: A novel
  approach},'' \emph{Applied Soft Computing Journal}, vol.~62, pp. 579--591,
  2018.

\bibitem{Antoniol2008}
\BIBentryALTinterwordspacing
G.~Antoniol, K.~Ayari, M.~{Di Penta}, F.~Khomh, and Y.-G.
  Gu{\'{e}}h{\'{e}}neuc, ``{Is it a bug or an enhancement?}'' in
  \emph{Conference of the center for advanced studies on collaborative research
  meeting of minds (CASCON '08)}.\hskip 1em plus 0.5em minus 0.4em\relax ACM
  Press, 2008, pp. 304----318. [Online]. Available:
  \url{http://portal.acm.org/citation.cfm?doid=1463788.1463819}
\BIBentrySTDinterwordspacing

\bibitem{Chawla2015}
\BIBentryALTinterwordspacing
I.~Chawla and S.~K. Singh, ``{An automated approach for bug categorization
  using fuzzy logic},'' in \emph{8th India Software Engineering Conference
  (ISEC)}.\hskip 1em plus 0.5em minus 0.4em\relax ACM, feb 2015, pp. 90--99.
  [Online]. Available:
  \url{http://dl.acm.org/citation.cfm?doid=2723742.2723751}
\BIBentrySTDinterwordspacing

\bibitem{Herzig2013}
K.~Herzig, S.~Just, and A.~Zeller, ``{It's not a bug, it's a feature: How
  misclassification impacts bug prediction},'' in \emph{International
  Conference on Software Engineering (ICSE 2013)}, 2013, pp. 392--401.

\bibitem{Lopes2020}
F.~Lopes, J.~Agnelo, C.~A. Teixeira, N.~Laranjeiro, and J.~Bernardino,
  ``{Automating orthogonal defect classification using machine learning
  algorithms},'' \emph{Future Generation Computer Systems}, vol. 102, pp.
  932--947, jan 2020.

\bibitem{Chawla2002}
\BIBentryALTinterwordspacing
N.~V. Chawla, K.~W. Bowyer, L.~O. Hall, and W.~P. Kegelmeyer, ``{SMOTE:
  Synthetic minority over-sampling technique},'' \emph{Journal of Artificial
  Intelligence Research}, vol.~16, pp. 321--357, jan 2002. [Online]. Available:
  \url{https://www.jair.org/index.php/jair/article/view/10302}
\BIBentrySTDinterwordspacing

\end{thebibliography}

\end{document}